\documentclass[prl,twocolumn,showpacs]{revtex4}

\usepackage{graphicx}
\usepackage{amsmath}

\begin{document}

\title{Correlations and Counting Statistics of an Atom Laser}

\author{Anton \"Ottl, Stephan Ritter, Michael K{\"o}hl$^\dag$, and Tilman Esslinger}

\affiliation{Institute of Quantum Electronics, ETH Z\"{u}rich,
H\"{o}nggerberg, CH--8093 Z\"{u}rich, Switzerland}
\date{\today}

\begin{abstract}
We demonstrate time-resolved counting of single atoms extracted
from a weakly interacting Bose-Einstein condensate of $^{87}$Rb
atoms. The atoms are detected with a high-finesse optical cavity
and single atom transits are identified. An atom laser beam is
formed by continuously output coupling atoms from the
Bose-Einstein condensate. We investigate the full counting
statistics of this beam and measure its second order correlation
function $g^{(2)}(\tau)$ in a Hanbury Brown and Twiss type
experiment. For the monoenergetic atom laser we observe a constant
correlation function $g^{(2)}(\tau)=1.00\pm0.01$ and an atom
number distribution close to a Poissonian statistics. A
pseudo-thermal atomic beam shows a bunching behavior and a Bose
distributed counting statistics.
\end{abstract}


\maketitle

Correlations between identical particles were first observed by
Hanbury Brown and Twiss in light beams \cite{HanburyBrown1956}.
Their idea was that intensity fluctuations and the resulting
correlations reveal information about the coherence and the
quantum statistics of the probed system. This principle has found
applications in many fields of physics \cite{Baym1998} such as
astronomy \cite{HanburyBrown1956b}, high-energy physics
\cite{Goldhaber1960}, atomic physics \cite{Shimizu1996} and
condensed matter physics \cite{Henny1999,Oliver1999}. In optics,
the reduced intensity fluctuations of a laser have been observed
by Arecchi \cite{Arecchi1965} only a few years after its
invention, thereby disclosing the extraordinary properties of this
light source.

With the realization of Bose-Einstein condensation in dilute
atomic gases a novel weakly interacting quantum system is
available. The interpretation of a Bose-Einstein condensate
representing a single, macroscopic wave function has been
supported in numerous experiments highlighting its phase coherence
\cite{Andrews1997,Stenger1999,Hagley1999,Bloch2000}.
Correspondingly, atom lasers are atomic beams which are coherently
extracted from Bose-Einstein condensates
\cite{Mewes1997,Anderson1998,Hagley1999b,Bloch1999}. Their first
order phase coherence has been observed both in space
\cite{Andrews1997} and time \cite{Koehl2001}. However, only the
second order coherence reveals whether atom lasers exhibit a truly
laser-like behavior. Here we present a measurement of the second
order correlation function $g^{(2)}(\tau)$ of an atom laser in a
Hanbury Brown and Twiss type experiment.

The second order correlation function $g^{(2)}(\tau)$ represents
the conditional likelihood for detecting a particle a time $\tau$
later than a previously detected particle and quantifies second
order coherence \cite{Glauber1963}. For a thermal source of bosons
$g^{(2)}(\tau)$ equals 2 for $\tau=0$ and decreases to 1 on the
time scale of the correlation time which is given by its energy
spread. For a coherent source $g^{(2)}(\tau)=1$ holds for all
times and therefore intensity fluctuations are reduced to the shot
noise limit. Higher order coherence in quantum degenerate samples
was so far only studied in the spatial domain where atom-atom
interactions reveal the short-distance correlations
\cite{Burt1997}. In an interferometric measurement $g^{(2)}(r)$
has been determined for elongated, phase-fluctuating condensates
\cite{Hellweg2003}, and recently spatial correlation effects in
expanding atom clouds were observed \cite{Foelling2005}.

\begin{figure}[htbp]
  \includegraphics[width=0.75\columnwidth]{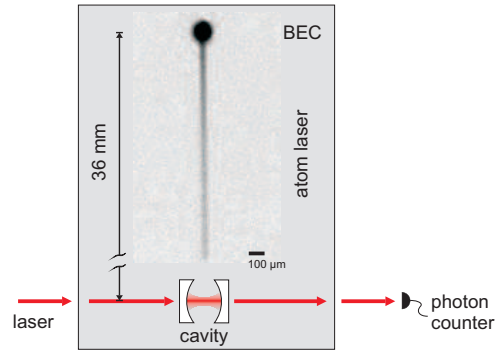}
  \caption{Schematic of the experimental setup. A weak continuous atom laser beam
  is released from a Bose-Einstein condensate. After dropping
  a distance of 36\,mm the atoms enter a high-finesse optical
  cavity and single atoms in the beam are detected. For the actual measurement the atomic flux
  is reduced by  factor $10^4$ as compared to the image.}
  \label{fig0}
\end{figure}

We demonstrate detection of single atoms from a weakly interacting
quantum gas by employing a high-finesse optical cavity
\cite{Mabuchi1996,Munstermann1999,Robert2001} (see fig.
\ref{fig0}). Detecting the arrival times of all atoms at the
cavity explicitly gives access to the full counting statistics
that reveals the atom number distribution function and its
statistical moments \cite{Levitov1996,Belzig2004}. Determining the
full counting statistics goes far beyond a measurement of the
intensity correlation function only, because it represents the
full statistical information about the quantum state. Despite
recent progress, especially in mesoscopic electronic systems
\cite{Reulet2003}, the full counting statistics has not been
measured for massive particles before. For neutral atoms this
quantity is of special interest, since the strength of the
interaction does not overwhelm the quantum statistics as it is
often the case for electrons.

Our new experimental design combines the techniques for the
production of atomic Bose-Einstein condensates with single atom
detection by means of a high-finesse optical cavity. The apparatus
consists of an ultra high vacuum (UHV) chamber which incorporates
a separated enclosure with a higher background pressure. Here we
collect $10^9$ $^{87}$Rb atoms in a vapor cell magneto-optical
trap which is loaded from a pulsed dispenser source. After
polarization gradient cooling and optical pumping into the $|F=1,
m_F=-1\rangle$ hyperfine ground state we magnetically transfer the
atoms over a distance of 8\,cm out of the enclosure into a
magnetic trap. All coils for the magnetic trapping fields are
placed inside the UHV chamber and are cooled to below 0$^\circ$C.
In the magnetic trap we perform radio frequency induced
evaporative cooling of the atomic cloud and obtain almost pure
Bose-Einstein condensates with $1.5\times 10^6$ atoms. After
evaporation we relax the confinement of the atoms to the final
trapping frequencies $\omega_\perp= 2\pi \times 29$\,Hz and
$\omega_\|=2 \pi \times 7$\,Hz, perpendicular and parallel to the
symmetry axis of the magnetic trap, respectively.

For output coupling an atom laser beam we apply a weak continuous
microwave  field to locally spin-flip atoms inside the
Bose-Einstein condensate into the $|F=2, m_F=0\rangle$ state.
These atoms do not experience the magnetic trapping potential but
are released from the trap and form a well collimated beam which
propagates downwards due to gravity \cite{Bloch1999}. The output
coupling is performed near the center of the Bose condensate for a
duration of 500\,ms during which we extract on the order of
$3\times 10^3$ atoms. After dropping a distance of 36\,mm the
atoms enter the high finesse optical cavity (see figure 1). Fine
tuning of the relative position between the atom laser beam and
the cavity mode is obtained by tilting the vacuum chamber. We
maintain a magnetic field along the trajectory of the atom laser,
which at the position of the cavity is oriented vertically and has
a strength of approximately 15\,G.

The cavity consists of two identical mirrors separated by
178\,$\mu$m. Their radius of curvature is 77.5\,mm resulting in a
Gaussian TEM$_{00}$ mode with a waist of $w_0=26\,\mu$m. The
coupling strength between a single Rb atom and the cavity field is
$g_0=2 \pi \times 10.4$\,MHz on the $F=2 \rightarrow F^\prime=3$
transition of the D$_2$-line. The cavity has a finesse of $3\times
10^5$ and the decay rate of the cavity field is $\kappa= 2 \pi
\times 1.4$\,MHz. The spontaneous emission rate of the rubidium
atom is $\Gamma= 2 \pi \times 6$\,MHz. Therefore we operate in the
strong coupling regime of cavity QED. The cavity mirrors are
mounted inside a piezo tube which enables precise mechanical
control over the length of the resonator \cite{Munstermann1999}.
Four radial holes in the piezo element allow atoms to enter the
cavity volume and also provide optical access perpendicular to the
cavity axis. The cavity resides on top of a vibration isolation
mount which ensures excellent passive stability. The cavity
resonance frequency is stabilized  by means of a far-detuned laser
with a wavelength of 831\,nm using a Pound-Drever-Hall locking
scheme.

\begin{figure}[htbp]
  \includegraphics[width=0.65\columnwidth]{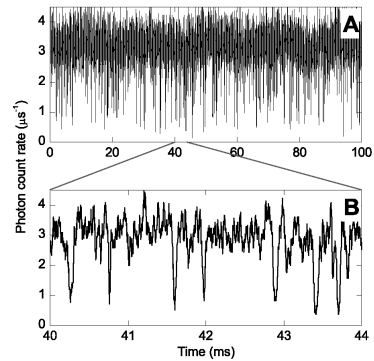}
  \caption{{\bf A:} Light transmission through the high-finesse optical cavity when an atom laser beam is
  transversing.  {\bf B:} Details of the single atom transits. The photon count rate is averaged over 20\,$\mu$s.}
  \label{fig1}
\end{figure}

The cavity is probed by a weak, near resonant laser beam, whose
transmission is monitored by a single photon counting module. The
presence of an atom inside the cavity results in a drop of the
transmission (see figure 2). The stabilization light is blocked
from the single photon counter by means of optical filters with an
extinction of 120\,dB. The probe laser and the cavity are
red-detuned from the atomic $F=2 \rightarrow F^\prime=3$
transition by 40\,MHz and 41\,MHz, respectively. The polarization
of the laser is aligned horizontally and the average intra-cavity
photon number is 5. These settings are optimized to yield a
maximum detection efficiency for the released atoms which is
$(23\pm 5)\%$. This number is primarily limited by the size of the
atom laser beam which exceeds the cavity mode cross section. The
atoms enter the cavity with a velocity of 84\,cm/s giving rise to
an interaction time with the cavity mode of typically 40\,$\mu$s,
which determines the dead-time of our detector. The dead-time of
our detector is short compared to the time scale of the
correlations, which allows us to perform Hanbury Brown and Twiss
type measurements with a single detector \cite{Mandel1965}.

We record the cavity transmission for the period of the atom laser
operation and average the photon counting data over 20$\mu$s (see
figure 2). Using a peak detection routine we determine the arrival
time of an atom in the cavity, requiring that the cavity
transmission drops below its background value by at least four
times the standard deviation of the photon shot noise. From the
arrival times of all atoms we compute the second order correlation
function $g^{(2)}(\tau)$ by generating a histogram of all time
differences within a single trace and normalizing it by the mean
atomic flux. Due to the finite duration of the measurement $T$,
the number of events with a time difference $\tau$ is reduced
according to $1-\tau/T$, which is taken into account by dividing
the correlation function by this factor. We average these
histograms over many repetitions of the experiment to obtain
$g^{(2)}(\tau)$ with a high signal-to-noise ratio.

\begin{figure}[htbp]
  \includegraphics[width=.72\columnwidth,clip=true]{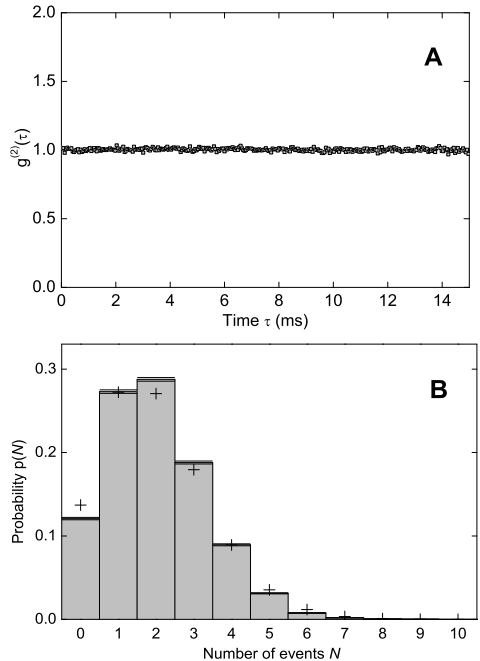}
  \caption{{\bf A:} Second order correlation function of an atom laser beam. The data are binned with a time bin size of 50\,$\mu$s.
  The average count number is $2\times10^5$ per bin. We have omitted the first two data points since they are
  modified by the dead-time of our detector.
  {\bf B:} Probability distribution p({\it N}) of the atom number {\it N} detected within a time interval of $T=1.5$\,ms.
  The (+) symbols show a Poissonian distribution for the same mean value of $\langle n \rangle=1.99$ as the measured data.
  The errors indicate statistical errors.}
  \label{fig2}
\end{figure}

Figure 3A shows the measured second order correlation function of
an atom laser beam. The value of the correlation function is
$g^{(2)}(\tau)=1.00\pm0.01$ which is expected for a coherent
source. The second order correlation function being equal to unity
reveals the second order coherence of the atom laser beam and is
intimately related to the property that it can be described by a
single wave function. Residual deviations from unity could arise
from technical imperfections. Magnetic field fluctuations either
due to current noise in the magnetic trapping coils or due to
external fluctuations could imprint small intensity fluctuations
onto the atom laser beam. We employ a low noise current source and
magnetic shielding to minimize these effects. In addition, we use
a highly stable microwave source which is stabilized to a GPS
disciplined oscillator. A further contribution to a potential
modification of the second order correlation function could be due
to the output coupling process itself. The spatial correlation
function of atoms output coupled from a weakly interacting
condensate has been studied theoretically in a situation
neglecting gravity \cite{Choi2000}. The modification from a
constant unity second order correlation function was on the order
of one percent, which is on the same order of magnitude as the
uncertainty in our data.

Measuring the second order correlation function requires to detect
the particles within their coherence time and coherence volume
\cite{Mandel1965}. The uncertainty of the detection time of an
atom must be smaller than the correlation time, because otherwise
the correlations vanish \cite{Shimizu1996}. We estimate that the
acquired time delays resulting from a possibly misaligned detector
are shorter than the dead time of our detector. It has been
measured that the coherence time of the atom laser is given by the
duration of output coupling at least for durations of 1.5\,ms
\cite{Koehl2001}.

Trapped Bose-Einstein condensates have been demonstrated to be
phase-coherent and to have a uniform spatial and temporal phase
\cite{Stenger1999,Hagley1999,Bloch2000}. The atom laser beam has
been theoretically described by a single wave function
\cite{Choi2000,Gerbier2001} and its spatial coherence was observed
\cite{Andrews1997}. Moreover, a full contrast interference pattern
was observed between two atom laser beams extracted from separate
locations inside a condensate \cite{Bloch2000}. This indicates a
high degree of spatial overlap between the two propagating modes
and a negligible distortion of the uniform spatial phase due to
interactions with the remaining condensate. From this we conclude
that the atom laser leaves the condensate region with a well
defined spatial wavefront.

Many overlapping spatial modes at the detector wash out the
correlations. In our experimental geometry this is the case when
output coupling from a thermal source, since we can not resolve a
single diffraction limited spatial mode. Therefore we do not
observe thermal bunching of non-condensed atoms.

Determining the arrival times of all detected atoms explicitly
allows us to extract the full counting statistics of the atoms. We
choose a time bin length of $T$=1.5\,ms in which we count the
number $N$ of detected atoms and plot the probability distribution
$p(N)$ (see fig. 3B). The distribution is close to a Poissonian
distribution $p(N)= \langle n \rangle ^N e^{-\langle n
\rangle}/N!$ with a mean of $\langle n \rangle=1.99$. For the
measured distribution we have calculated the 2nd, 3rd and 4th
cumulant to be $\kappa_2=1.75$, $\kappa_3=1.34$ and
$\kappa_4=0.69$, respectively. We attribute the small deviation
from the Poissonian distribution to having two or more atoms
arriving within the dead time of our detector. For the total flux
of 5.2 atoms per ms this probability is $5\%$.

\begin{figure}[htbp]
  \includegraphics[width=.72\columnwidth,clip=true]{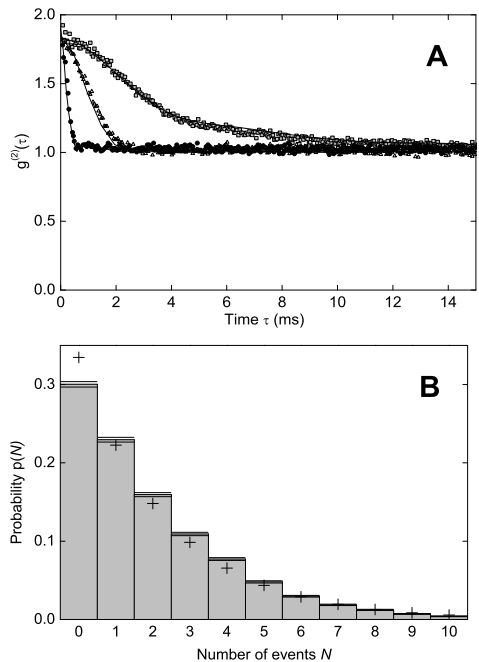}
  \caption{{\bf A:} Second order correlation functions of pseudo-thermal atomic beams. The square symbols
  correspond to a filter band width (FWHM) of 90\,Hz, the triangles to a bandwidth of 410\,Hz, and the circles to a bandwidth
  of 1870\,Hz. The data are binned with a time bin size of 50\,$\mu$s in which the average count number is $8\times10^4$.
  We have omitted the first two data points since they are
  modified by the dead-time of our detector. The lines are
  the experimentally determined correlation functions of the broadband microwave fields.
  {\bf B:} Probability distribution p({\it N}) of the atom number {\it N} within a time interval of $T$=1.5\,ms for the 90\,Hz bandwidth data.
  The (+) symbols indicate a Bose distribution with the same mean value of $\langle n \rangle=1.99$. The errors indicate the
  statistical errors.}
  \label{fig3}
\end{figure}

We realize a direct comparison with a pseudo-thermal beam of atoms
by output coupling a beam with thermal correlations from a
Bose-Einstein condensate. This is in close analogy to changing the
coherence properties of a laser beam by means of a rotating ground
glass disc \cite{Arecchi1965}. Instead of applying a monochromatic
microwave field for output coupling we have used a broadband
microwave field with inherent frequency and intensity noise. We
have employed a white noise generator in combination with  quartz
crystal band pass filters which set the bandwidth of the noise.
The filters operate at a frequency of a few MHz and the noise
signal is subsequently mixed to a fixed frequency signal at
6.8\,GHz to match the output coupling frequency. For atomic beams
prepared in such a way we observe bunching with a time constant
set by the band pass filter (see figure 4A). To compare our data
with the theoretically expected correlation function we have
measured the power spectra of the band pass filters and calculated
$|g^{(1)}(\tau)|^2$ of the rf field before frequency  mixing. In
figure 4A we plot $1+\beta |g^{(1)}(\tau)|^2$. The normalization
factor $\beta= 0.83$ accounts for the deviation of the
experimental data from $g^{(2)}(0)=2$ due to imperfections in the
frequency mixing process.

For the pseudo-thermal beam we also calculate the counting
statistics and find a significantly different behavior than for
the atom laser case. For a filter with a spectral width (FWHM) of
90\,Hz we have chosen the time bin length of $T$=1.5\,ms, smaller
than the correlation time. The atomic flux with a mean atom number
$\langle n \rangle=1.99$ is equal to the case of the atom laser.
We compare the measured probability distribution to a Bose
distribution $p(N)=\langle n \rangle^N/(1+\langle n
\rangle)^{1+N}$, which is expected for a thermal sample and find
good agreement (see figure 4B). From the distribution we have
extracted the 2nd, 3rd and 4th cumulant to be $\kappa_2=4.6$,
$\kappa_3=14.5$ and $\kappa_4=50.6$, respectively.

In conclusion we have demonstrated the detection of single atoms
from a quantum degenerate source with an efficiency of $23\%$ and
measured the second order correlation function of an atom laser.
We find the atom laser to be second order coherent. Moreover, the
counting statistics of the atom laser was measured and higher
moments of the number distribution were extracted.

We would like to thank W. Belzig, T. Bourdel, C. Bruder, T.
Donner, A. Frank, G. Morigi, H. Ritsch, T. St\"oferle, J.P. Stucki
for discussions and SEP Information Sciences, OLAQUI and QSIT for
funding.

\end{document}